\documentclass[lettersize,journal]{IEEEtran}
\usepackage{amsmath,amsfonts}
\usepackage{algorithmic}
\usepackage{algorithm}
\usepackage{array}
\usepackage[caption=false,font=normalsize,labelfont=sf,textfont=sf]{subfig}
\usepackage{textcomp}
\usepackage{stfloats}
\usepackage{url}
\usepackage{verbatim}
\usepackage{graphicx}
\usepackage{cite}
\usepackage{multirow}
\usepackage{bbding}
\usepackage{utfsym}
\usepackage{balance}
\DeclareUnicodeCharacter{2212}{\ensuremath{-}}
\begin{document}

\title{InstructTTS: Modelling Expressive TTS in Discrete Latent Space with Natural Language Style Prompt}
\author{Dongchao Yang{*}, Songxiang Liu{*}, Rongjie Huang, Chao Weng, Helen Meng,~\IEEEmembership{Fellow,~IEEE}
\thanks{Dongchao Yang and Helen Meng are with the Chinese University of Hong Kong. This work was done when Dongchao Yang was an intern at Tencent AI Lab. {*} denotes equal contribution with order determined by alphabetic order.}
\thanks{Songxiang Liu and Chao Weng are with Tencent AI Lab.}
\thanks{Rongjie Huang is with the Zhejiang University, China.}
\thanks{Songxiang Liu is the corresponding author.}}

\markboth{Journal of \LaTeX\ Class Files,~Vol.~14, No.~8, August~2021}%
{Shell \MakeLowercase{\textit{et al.}}: A Sample Article Using IEEEtran.cls for IEEE Journals}


\maketitle
\begin{abstract} Expressive text-to-speech (TTS) aims to synthesize speech with varying speaking styles to better reflect human speech patterns. 
In this study, we attempt to use natural language as a style prompt to control the styles in the synthetic speech, \textit{e.g.}, ``Sigh tone in full of sad mood with some helpless feeling".
Considering that there is no existing TTS corpus that is suitable to benchmark this novel task, we first construct a speech corpus whose speech samples are annotated with not only content transcriptions but also style descriptions in natural language.
Then we propose an expressive TTS model, named InstructTTS, which is novel in the sense of the following aspects:
(1) We fully take advantage of self-supervised learning and cross-modal metric learning and propose a novel three-stage training procedure to obtain a robust sentence embedding model that can effectively capture semantic information from the style prompts and control the speaking style in the generated speech.
(2) We propose to model acoustic features in discrete latent space and train a novel discrete diffusion probabilistic model to generate vector-quantized (VQ) acoustic tokens rather than the commonly-used mel spectrogram.
(3) We jointly apply mutual information (MI) estimation and minimization during acoustic model training to minimize style-speaker and style-content MI, avoiding possible content and speaker information leakage from the style prompt.
Extensive objective and subjective evaluation has been conducted to verify the effectiveness and expressiveness of InstructTTS. Experimental results show that InstructTTS can synthesize high-fidelity and natural speech with style prompts controlling the speaking style. Synthesized samples are available online \footnote{http://dongchaoyang.top/InstructTTS/}.

\end{abstract}

\section{Introduction}
\IEEEPARstart{T}{ext}-to-speech (TTS) aims to generate human-like speech from input text, which attracts broad interest in the audio and speech processing community. Nowadays, the state-of-the-art TTS systems \cite{wang2017tacotron, ren2020fastspeech, kim2021conditional} are able to produce natural and high-quality speech.
However, there still exists a big gap between TTS-synthetic speech and human speech in terms of expressiveness, which limits the broad applications of current speech synthesis systems.
Many researchers now focus on a more challenging task, i.e., expressive TTS, which aims to model and control the speaking style (\textit{e.g.}, emotion, speaking-rate and so on) in the generated speech according to human's demands.
We note that there are generally two types of methods in the literature to learn speaking style information: one type uses auxiliary categorical style labels as the condition of the framework \cite{tits2019visualization,tits2019exploring}, the other
imitates the speaking style of a reference speech \cite{wang2018style,skerry2018towards,jia2018transfer,yang2022norespeech}.
However, there are limitations in the diversity of expressiveness when categorical style labels are used, as these models can only generate a few pre-defined styles from the training set.
While TTS models that use a reference utterance to generate a particular speaking style can be trained in an unsupervised manner and are generalizable to out-of-domain speaking styles, the style information extracted from the reference speech may not be easily understandable or interpretable. Additionally, it can be challenging to select a reference speech sample that precisely matches a user's requirements.

For the first time, we study the modelling of expressive TTS with style prompt in natural language, where we meet with the following research problems: (1) how to train a language model that can capture semantic information from the natural language prompt and control the speaking style in the generated speech; (2) how to design an acoustic model to effectively model the challenging one-to-many learning problem of expressive TTS. In this paper, we will address these two challenges.

The main contributions of this study are summarized as follows: \\
(1) For the first time, we study the modelling of expressive TTS with natural language prompts, which brings us a step closer to achieving user-controllable expressive TTS. \\
(2) We introduce a novel three stage training strategy to obtain a robust sentence embedding model that can effectively capture semantic information from the style prompts. \\
(3) We propose to model acoustic features in discrete latent space and cast speech synthesis as a sequence-to-sequence language modeling task. Specifically, we train a novel discrete diffusion model to generate vector-quantized (VQ) acoustic features rather than to predict the commonly-used hand-crafted intermediate acoustic features, such as the mel-spectrogram. \\
(4) We explore to model two types of VQ acoustic features: mel-spectrogram based VQ features and waveform-based VQ features. We demonstrate that the two types of VQ features can be effectively modeled by our proposed novel discrete diffusion model. Our waveform-based modelling method only needs one-stage training, and it is a non-autoregressive model, which is far different from the concurrent work VALL-E \cite{wang2023neural} and MusicLM \cite{borsos2023musiclm}. \\
(5) We jointly apply mutual information (MI) estimation and minimization during acoustic model training to minimize style-speaker and style-content MI, avoiding possible content and speaker information leakage from the style prompt.


The rest of this paper is organized as follows: In Section~\ref{sec2:intro}, we motivate our study by introducing the background and related work. In Section~\ref{sec3:dataset}, we present the details of the dataset. In Section~\ref{sec4:Proposed method}, we introduce the details of our proposed methods. The experimental setting, evaluation metrics and results are presented from Section~\ref{sec5:exp} to Section~\ref{sec7:result}. The study is concluded in Section~\ref{sec8:conclusion}.

\section{Related work and background}
\label{sec2:intro}
This study is built on several previous works on cross-modal representation learning, vector quantization, diffusion probabilistic models and expressive TTS. We briefly introduce the related studies to set the stage for our research and rationalize the novelty of our contributions.
\begin{figure*}[t] 
  \centering
  \includegraphics[width=\linewidth]{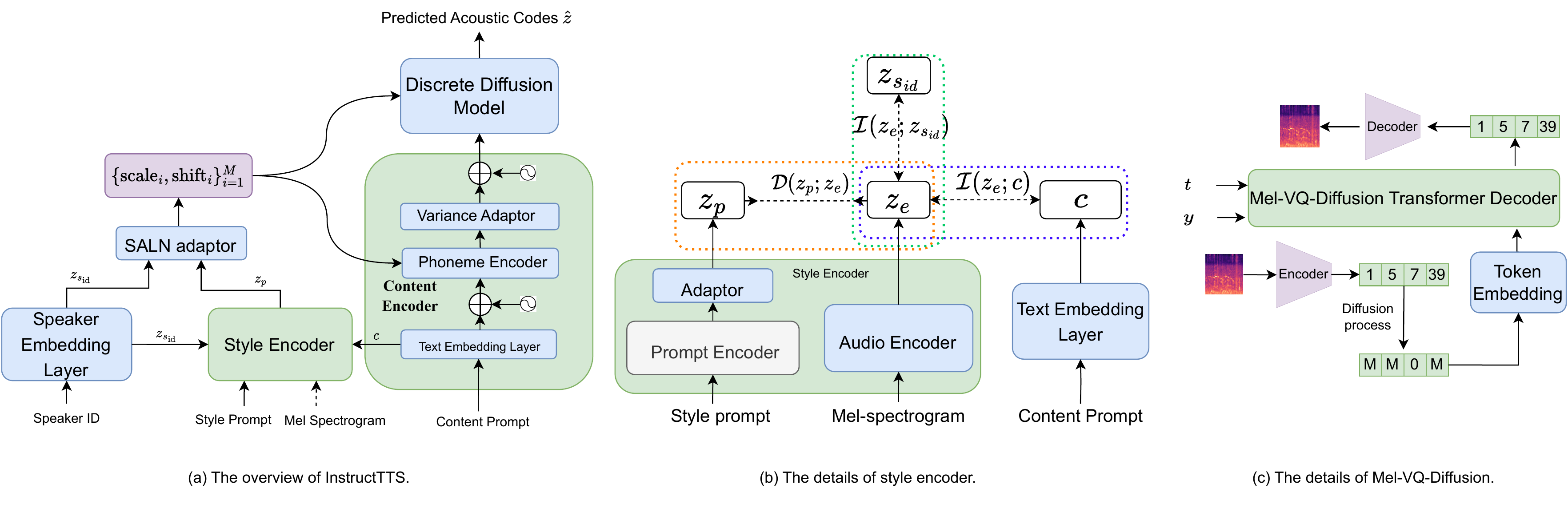}
  \caption{(a) shows the model architecture of our proposed InstructTTS. Where SALN denotes the style-adaptive layer normalization adaptor \cite{min2021meta}. (b) shows the details of our proposed style encoder, which aims to extract style features from GT mel-spectrogram (training stage) or style prompt (inference stage). In Figure 1 (c), we give an example of discrete diffusion decoder to generate VQ mel-spectrogram acoustic features (we name it as Mel-VQ-Diffusion).}
  \label{fig:1}
  \vspace*{-\baselineskip}
\end{figure*}

\subsection{Cross-modal Representation Learning} 
 Cross-modal representation learning aims to learn a common latent space for different modal data (\textit{e.g.} text and image, text and speech). In general, two different modal encoders are used to extract deep feature representation, and then a variety of supervised or unsupervised strategies are devised to align the two modal representation spaces \cite{radford2021learning,koepke2022audio}. In our study, we expect to control the acoustic features (such as pitch, emotion and speed) by a natural language sentence. To realize this target, we turn to cross-modal representation learning. The details will be discussed later.

\subsection{Vector Quantization}
Vector quantization technique has been widely used in different fields, such as image \cite{van2017neural,razavi2019generating,esser2021taming} and speech processing \cite{baevski2019vq,hsu2021hubert,defossez2022high,zeghidour2021soundstream}. VQ-VAE \cite{van2017neural} is one of the most successful vector quantization models. VQ-VAE uses an encoder and a quantizer to compress an image into a low-dimensional discrete space, then a decoder is used to reconstruct the image from a sequence of discrete tokens. Inspired by VQ-VAE, a series of works adopt the idea to reconstruct mel-spectrogram or linear-spectrogram \cite{iashin2021taming,yang2022diffsound}. Recently, many works have focused on reconstruct waveform by VQ-VAE. To supplement the information loss during the VQ process, Residual-VQ (RVQ) \cite{zeghidour2021soundstream} and Group-residual-VQ (GRVQ) \cite{yang2023hifi} technique are proposed, which uses multiple different codebooks to encode the audio information. Nowadays, the majority of TTS systems focus on using an acoustic model (AM) to directly predict mel-spectrogram, then uses a pre-trained vocoder to recover waveform from the predicted mel-spectrogram \cite{ren2020fastspeech,elias2021parallel,kim2020glow}. However, the mel-spectrogram is highly correlated along both time and frequency axes in a complicated way, leading to a great difficulty for the AM to predict. Furthermore, the gap between the ground-truth (GT) mel-spectrogram and the predict one from AM degrades the performance due to the vocoder is trained on GT mel-spectrogram. 
In this study, instead of using AM to predict mel-spectrogram, we turn to predict learnable and vector-quantized acoustic representation, which is transformed to a discrete latent space.

\subsection{Expressive Text-to-speech} 
In the field of expressive speech synthesis, several works \cite{wang2018style,li2021towards,min2021meta,casanova2021sc,zhou2022speech,huang2022generspeech,yang2022norespeech} have been done. Wang \textit{et al.} \cite{wang2018style} propose to use global style tokens to control and transfer global speaking styles. Inspired by \cite{wang2018style}, many similar works propose to learn speaking style from a reference audio, such as, Li \textit{et al.} \cite{li2021towards} use a multi-scale style encoder capturing style information from reference audio to facilitate expressive speech synthesis; Huang \textit{et al.} \cite{huang2022generspeech} propose a multi-level style adaptor to transfer speaking style.
The most related to our work are Style-Tagging-TTS (ST-TTS) \cite{kim2021expressive} and PromptTTS \cite{guo2022prompttts}. ST-TTS proposes to use style tags to guide the speaking style of synthsized speech, where style tags denote a short phrase or word representing the style of an utterance, such as emotion, intention, and tone of voice. In this study, we attempt to use longer natural language as style descriptions to control the styles in the synthetic speech, which is more complicated due to the fact that longer natural language prompts carry out more abundant semantic information and result in more complicated acoustic characteristics. Our concurrent work PromptTTS \cite{guo2022prompttts} proposed a similar idea with us, using a sentence as a style prompt to control the style information in TTS systems. They define five different style factors (gender, pitch, speaking speed, volume, and emotion), and assume the prompts have obvious style factor words, such as low-pitch, high-speaking speech and so on, which means that the model can obtain style information from local-level description.
Different from PromptTTS, our study does not apply constraint on the form of the style prompts and allows the user to use any free-form natural language to describe a speaking style, resulting in a much more challenging machine learning problem. Furthermore, we focus on Mandarin Chinese TTS and construct the first Mandarin Chinese speech corpus applicable for style-prompt-controllable expressive TTS.

\subsection{Diffusion Probabilistic Models}
Diffusion models have been demonstrated as a strong generative model for image generation \cite{dhariwal2021diffusion,gu2021vector,nichol2021glide,esser2021imagebart}, text generation \cite{li2022diffusion, gong2022diffuseq} and audio generation \cite{kong2020diffwave,jeong2021diff,yang2022diffsound,huang2023make}.
Roughly speaking, Diffusion models can be divided into two types according to whether the latent spaces are continuous or discrete. Nowadays, most diffusion models focus on continuous latent space, and the Latent Diffusion Model \cite{rombach2022high} is one of the representative works. Diffusion models with discrete state spaces are first investigated by \cite{sohl2015deep, hoogeboom2021argmax,austin2021structured}. These works try to define a structured categorical corruption process to corrupt the forward process, then train a model to learn the reverse process. Many works have successfully applied discrete diffusion models in image or sound generation, \textit{e.g.}, D3PMs \cite{austin2021structured} VQ-Diffusion \cite{gu2021vector}, DiffSound \cite{yang2022diffsound}. However, no one attempts to apply the discrete diffusion model to synthesize expressive human speech. In the following, we briefly review background knowledge of diffusion models.

\subsubsection{Vanilla Diffusion Model}
Diffusion models define a Markov chain $q(\boldsymbol{x}_{1:T}|\boldsymbol{x}_0)=\prod_{t=1}^{T}q(\boldsymbol{x}_t|\boldsymbol{x}_{t-1})$ that gradually destroys
data $x_0$ by adding noise over a fixed number of steps T,
so that $\boldsymbol{x}_T$ belongs to special noise distribution (\textit{e.g.} Gaussian distribution). The reverse process using a generative model that gradually denoises towards the original data distribution $p(\boldsymbol{x}_0)$. 
\subsubsection{Discrete Diffusion Model} \label{bkg:ddm}
Discrete diffusion models constrain variable $\boldsymbol{x}_t$ in  the state space so that  $\boldsymbol{x}_t$ is a discrete random variable falling into one of K categories. For discrete diffusion, a transition probability matrix is defined to indicate how $\boldsymbol{x}_0$ transits to $\boldsymbol{x}_t$. The forward process can be represented as categorical distributions $q(\boldsymbol{x}_t|\boldsymbol{x}_{t−1}) = \mathit{Cat}(\boldsymbol{x}_t; \boldsymbol{p}=\boldsymbol{x}_{t-1}\boldsymbol{Q}_t)$ where $\boldsymbol{Q}_t$
denotes the probabilities transition matrix. The more details about discrete diffusion models, please refer to \cite{yang2022diffsound,gu2021vector,austin2021structured}.
\begin{table*}[t] \centering
\caption{Example style prompts from different corpora. Since NLSpeech corpus is in Mandarin Chinese, we provide the translated version. The FSNR0 is in Korean, we provide the translated ones in the table.}
\label{tab:my-table1}
\begin{tabular}{ccc}
\hline
FSNR0 \cite{kim2021expressive}  & PromptSpeech \cite{guo2022prompttts}                                                                                                                                                   & NLSpeech (translated)                                                                                                                                                      \\ \hline
Seem sad   & \begin{tabular}[c]{@{}c@{}}A distressful male  sound appeared in low volume\end{tabular}                                          & \begin{tabular}[c]{@{}c@{}}The tone of the shock question revealed the sad feelings.\end{tabular}                                                                        \\ \hline
Bitter     & \begin{tabular}[c]{@{}c@{}}He sadly turns down  his volume, pitch and speed\end{tabular}              & \begin{tabular}[c]{@{}c@{}}It was a fiery expression of disapproval and condemnation, \\ with a palpable sense of irony,  a tinge of disgust and disdain.\end{tabular} \\ \hline
Pleased    & \begin{tabular}[c]{@{}c@{}}The ladylike person made an  increment of the volume and pitch\end{tabular}         & \begin{tabular}[c]{@{}c@{}}There was a sense of joy in the words, an expression of \\ joy in the heart, mixed with pride.\end{tabular}                                  \\ \hline
In a hurry & \begin{tabular}[c]{@{}c@{}}Men, low tone, said loudly and quickly\end{tabular}                                             & \begin{tabular}[c]{@{}c@{}}His voice grew more agitated, \\ and his tone revealed an urge and urgency.\end{tabular}                                                        \\ \hline
\end{tabular}
\end{table*}
\section{DATASET}
\label{sec3:dataset}
We use an internally collected Mandarin Chinese speech corpus named \textbf{NLSpeech} to conduct experimental evaluation since there is no openly available Mandarin Chinese speech corpus with rich style prompts. The corpus contains 44 hours of speech data (in total 32k utterances) from 7 speakers (5 female and 2 male). 
Audio waveform has a sampling rate of 24kHz. We randomly spare 0.1 hours of data as the validation set, another 0.1 hours of data as the test set and the remaining data as the training set. 
Each utterance has 5 style prompts labeled by different annotators. To obtain high-quality annotations, we ask annotators to follow three steps of annotation strategy: 
\begin{itemize}
    \item Step-1: The annotators first use one word to describe the overall perceived emotion of an utterance;
    \item Step-2: The annotators then listen to the utterance carefully and describe the emotion level of the utterance with one word;
    \item Step-3: The annotators write a complete sentence in natural language to describe the style of the utterance. 
\end{itemize}
Note that we ask annotators to not care about the speech content, which may influence the perception of emotion and style. Table \ref{tab:my-table1} shows example style prompts in our dataset, and we also compare NLSpeech with other existing related corpora, including the FSNR0 corpus \cite{kim2021expressive} and the PromptSpeech corpus \cite{guo2022prompttts}.
We note that the style prompts in NLSpeech are in free-form natural language sentences which are more consistent with those used in our daily life, while those in the FSNR0 and the PromptSpeech corpora are in controlled format. 
Meanwhile, this also brings us a challenging TTS problem since natural language sentences allow for expressing virtually any concepts. Compact and informative representation of style prompt is therefore paramount to achieve effective style controlling during speech synthesis.

\begin{figure}[t] \label{fig:cross-modal}
  \centering
  \includegraphics[width=\linewidth]{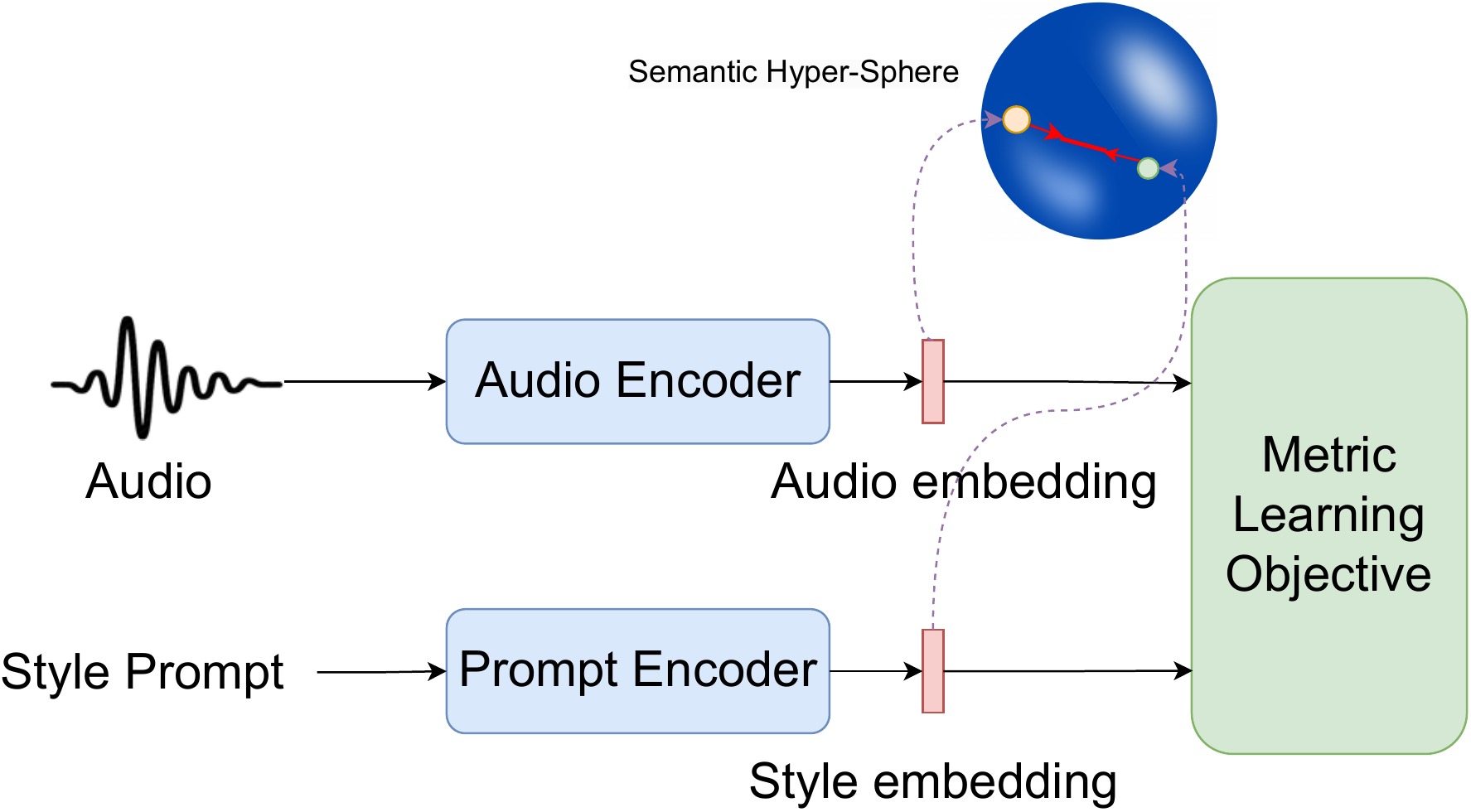}
  \caption{The model architecture of cross-modal representation learning.}
  \label{fig:2}
  \vspace*{-\baselineskip}
\end{figure}

\section{Proposed Method} \label{sec4:Proposed method}
The overall architecture of the proposed InstructTTS framework is demonstrated in Figure \ref{fig:1}, which consists of five parts, including a content encoder, a style encoder, a speaker embedding module, a style-adaptive layer normalization (SALN) adaptor and a discrete diffusion decoder. The detailed design of each part will be introduced in this section.

\subsection{Content Encoder} \label{text encoder}
The content encoder aims to extract content representation from the content prompts. We follow the architecture of FastSpeech2 \cite{ren2020fastspeech}, which consists of 4 feed-forward transformer. The hidden size, number of attention heads,
kernel size and filter size of the one-dimensional convolution
in the FFT block are set as 256, 2, 9 and 1024, respectively. After that, a variance adaptor is used to predict information such as duration and pitch that is closely related to the style of synthetic speech.


\subsection{Style Prompt Embedding Model} \label{Prompt Embedding}
To extract style representation from the style prompts, we adopt the RoBERTa model \cite{liu2019roberta} as prompt embedding model. Assuming we have a style prompt sequence $S=[S_1, S_2, ..., S_M]$, where $M$ denotes the sequence length. We add a [CLS] token to the start of the prompt sequence and then feed it into the prompt embedding model. After that, we take the representation of the [CLS] token as the style representation of this sentence.
In order to stably control the style of the output of TTS through natural language description, the quality of prompt embedding is of great importance, which should satisfy two conditions: (1) the learned style prompt space must be able to contain important semantic information; (2) the distribution of prompt embedding space should be relatively uniform and smooth, and the model can be generalized to the style description not seen in the training. To realize this target, we propose a novel three-stage training-fine-tuning strategy. The details are shown as follows.
\subsubsection{Training a base language model for Chinese}
Given that most open-source pre-trained language models are trained on English data, we first train a RoBERTa model on Chinese data. 
\subsubsection{Fine-tuning the pre-trained language model on labeled data}
We use a small amount of Chinese natural language inference (NLI) to fine-tune the model parameters in a supervised way to achieve a better semantic representation of the model. Specifically, we follow the training strategy proposed in SimCSE \cite{gao2021simcse}, which using an InfoNCE loss \cite{oord2018representation} objective to fine-tune our pre-trained RoBERTa model.
\subsubsection{Cross-modal representation learning between style prompts and speech}
We hope that the prompt embedding vector from the style prompt sentence and the style representation vector from the speech can be mapped to the shared semantic space so that we can control the style in the TTS output through the style description in the inference stage. Thus, we propose a cross-modal representation learning process based on metric learning, as Figure \ref{fig:cross-modal} shows. Specifically, we build an audio-text retrieval task based on the style-prompt and audio pair in our NLSpeech dataset. For any style prompt, we randomly choose $N-1$ negative audio samples, combined with one positive audio sample to build a training batch. Similarly, for one audio sample, we can also build a training batch that includes one positive style prompt and $N-1$ negative style prompts. Inspired by previous audio-text retrieval works \cite{koepke2022audio,chao20223cmlf}, we try to use contrastive ranking loss \cite{chopra2005learning} and InfoNCE loss \cite{oord2018representation} as the training objective. Experiments results show that InfoNCE loss brings better retrieval performance. The details will be introduced in Experiments part. 
\subsection{Style encoder}
The style encoder module, as shown in Fig.~\ref{fig:1} (b), includes three parts: A pre-trained robust style prompt embedding model (i.e., the prompt encoder), an adaptor layer to map the style embedding extracted from the prompt encoder into a new latent space, an audio encoder that extracts style information from the reference mel-spectrogram. Note that our pre-trained robust prompt embedding model is fixed when we train the remaining parts in the TTS model.
In the training stage, one of the training targets is to minimize the distance between style prompt embedding and audio embedding. We note that the audio encoder may encode speaker and content information. To make sure the audio encoder only encodes style-related information, during training, we jointly minimize the style-speaker mutual information (i.e., $I(\boldsymbol{z}_e;\boldsymbol{z}_{sid})$) and style-content mutual information (i.e., $I(\boldsymbol{z}_e;c)$). 
Mutual information (MI) is a key measurement of correlation between random variables. However, the MI of high-dimensional random variables with an unknown distribution is intractable to compute. Previous works focused on either estimating the MI lower bound or the MI upper bound. MINE \cite{pmlr-v80-belghazi18a} and InfoNCE \cite{van2017neural} compute a lower bound as the MI estimators while CLUB \cite{cheng2020club} computes an upper bound as the MI estimator. In this work, we use the CLUB method to minimize style-speaker and style-content MI to avoid content and speaker information leakage from the mel-spectrogram during training.
\begin{figure}[t] 
  \centering
  \includegraphics[width=\linewidth]{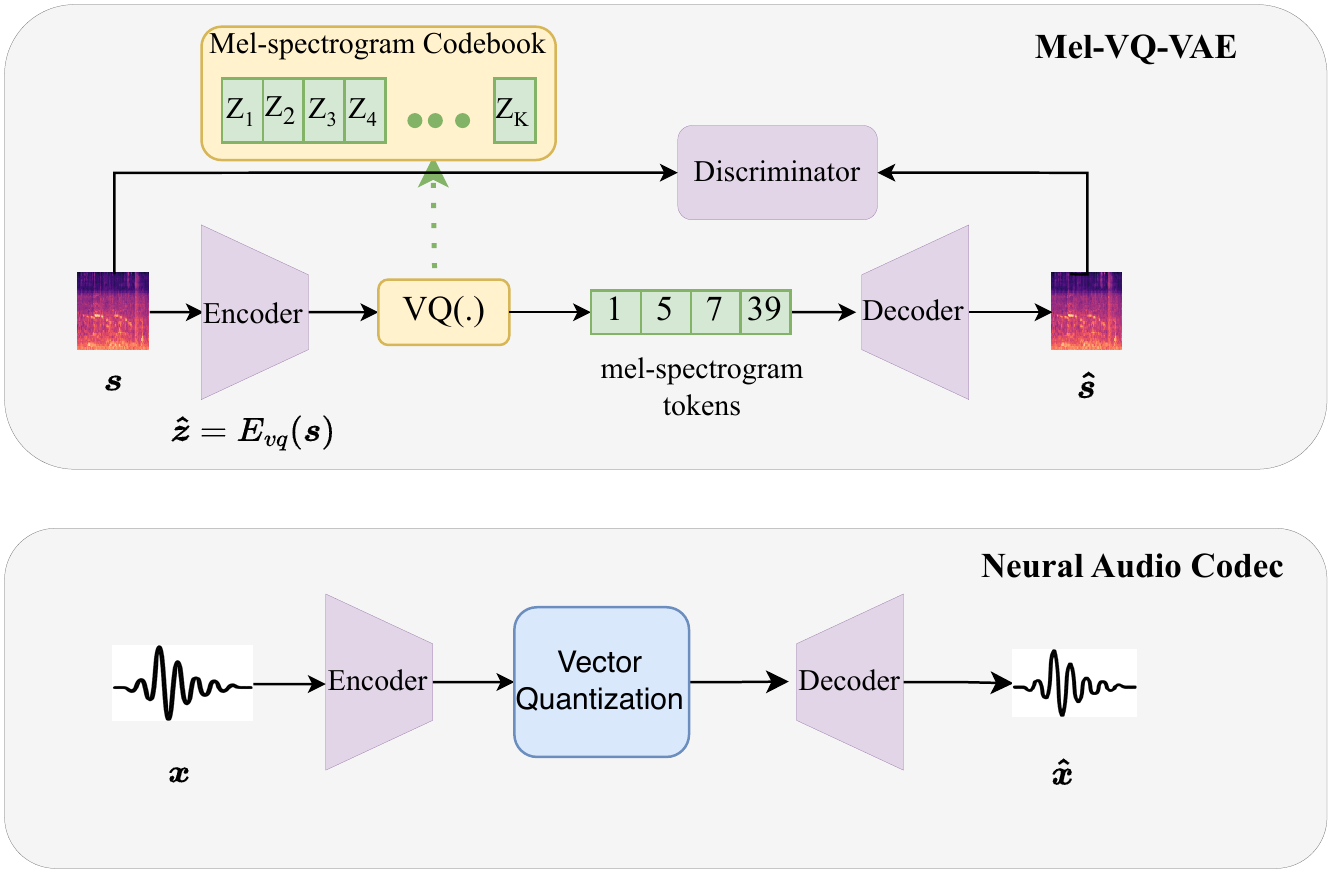}
  \caption{The overall architecture of the Mel-VQ-VAE and Neural Audio Codec Models.}
  \label{fig:vq-codec}
  \vspace*{-\baselineskip}
\end{figure}


\subsection{Modelling Mel-spectrograms in Discrete Latent Space}
In this part, we introduce our hypothesis: modelling mel-spectrograms in discrete latent space is a suitable way for expressive TTS. Then we introduce how to utilize VQ-VAE as an intermediate representation to help model the mel-spectrogram. Lastly, we introduce our proposed non-autoregressive mel-spectrogram token generation model, which is based on discrete diffusion models.

Most text-to-speech (TTS) methods \cite{tan2021survey,ren2020fastspeech,jeong2021diff} directly learn the mapping from text to mel-spectrogram in continuous space.
Then they use a pre-trained vocoder to decode the predicted mel-spectrogram into waveform. However, frequency bins in a mel-spectrogram are highly correlated along both time and frequency axes in a complicated way, especially when the speech sample conveys highly expressive emotions and speaking styles, leading to a challenging modeling problem.
Furthermore, the gap between the ground-truth mel-spectrogram and the predicted one also influences the synthesis performance~\cite{kong2020hifi}. In this study, we propose to model the mel-spectrogram in discrete latent space, but still use a HiFi-GAN vocoder \cite{kong2020hifi} to recover the waveform from the mel-spectrogram.
Specifically, we first pre-train a VQ-VAE with a large-scale speech dataset so that the pre-trained Mel-VQ-VAE encodes all of the linguistic, pitch, energy, and emotion information into the latent codes. Then we regard the vector quantized latent codes as the predicting targets and hence model the mel-spectrogram in the discrete latent space.
A similar idea modeling mel-spectrogram in discrete latent space is applied in VQ-TTS \cite{du2022vqtts}, which utilizes self-supervised
VQ acoustic feature (vq-wav2vec \cite{baevski2019vq}) rather than traditional mel-spectrogram as intermediate prediction target. VQ-TTS builds an autoregressive classification model for prosody labels and VQ acoustic features. Different from VQ-TTS, we still use the mel-spectrogram as intermediate acoustic feature and use a Mel-VQ-VAE model to transform the mel-spectrogram into a latent discrete space for reducing the time-frequency correlations.
As Figure \ref{fig:vq-codec} shows, a mel-spectrogram can be represented by a sequence of mel-spectrogram tokens. Thus, the mel-spectrogram synthesis problem transfers to predicting a sequence of discrete tokens, which can be seen as a language modeling problem. In the following, we will introduce the details of Mel-VQ-VAE, and then we will introduce our proposed Mel-VQ-Diffusion decoder.
 \subsubsection{Mel-VQ-VAE}
As Figure \ref{fig:vq-codec} shows, Mel-VQ-VAE has three parts: an Mel-encoder $E_{mel}$, Mel-decoder $G_{mel}$ and a codebook $\boldsymbol{Z}=\{ \boldsymbol{z}_k\}^K_{k=1} \in \mathbb{R}^{K \times n_z}$. we assume that the size of the codebook is $K$ and the dimension of codes is $n_z$. Assuming input a mel-spectrogram $\boldsymbol{s} \in \mathbb{R}^{F_{bin} \times T_{bin}}$, the mel-spectrogram is firstly encoded as a latent representation  $\boldsymbol{\hat{z}}=E_{mel}(\boldsymbol{s}) \in \mathbb{R}^{F^{\prime}_{bin} \times T^{\prime}_{bin} \times n_z}$ where $F^{\prime}_{bin} \times T^{\prime}_{bin}$ represents the reduced frequency and time dimension . Then we use the quantizer $Q(.)$ to map each feature $\boldsymbol{\hat{z}}_{ij}$ into its closest codebook entry $\boldsymbol{z}_k$ to obtain a sequence of discrete spectrogram tokens $\boldsymbol{z}_q$
\begin{equation}\label{codebook}
    \boldsymbol{z}_{q} = Q(\boldsymbol{\hat{z}}) := \big{(}\mathop{\arg\min}\limits_{\boldsymbol{z}_k \in \boldsymbol{Z}}||\boldsymbol{\hat{z}}_{ij}-\boldsymbol{ z}_k||_2^2 \big{)}
\end{equation}
Lastly, we use the decoder to reconstruct the mel-spectrogram, \textit{i.e.}, $\hat{\boldsymbol{s}}=G_{mel}(\boldsymbol{z}_{q})$. To improve the reconstruction performance, we follow VQGAN \cite{esser2021taming} adds an adversarial loss \cite{isola2017image} in the training stage.

\begin{figure}[t] 
  \centering
  \includegraphics[width=\linewidth]{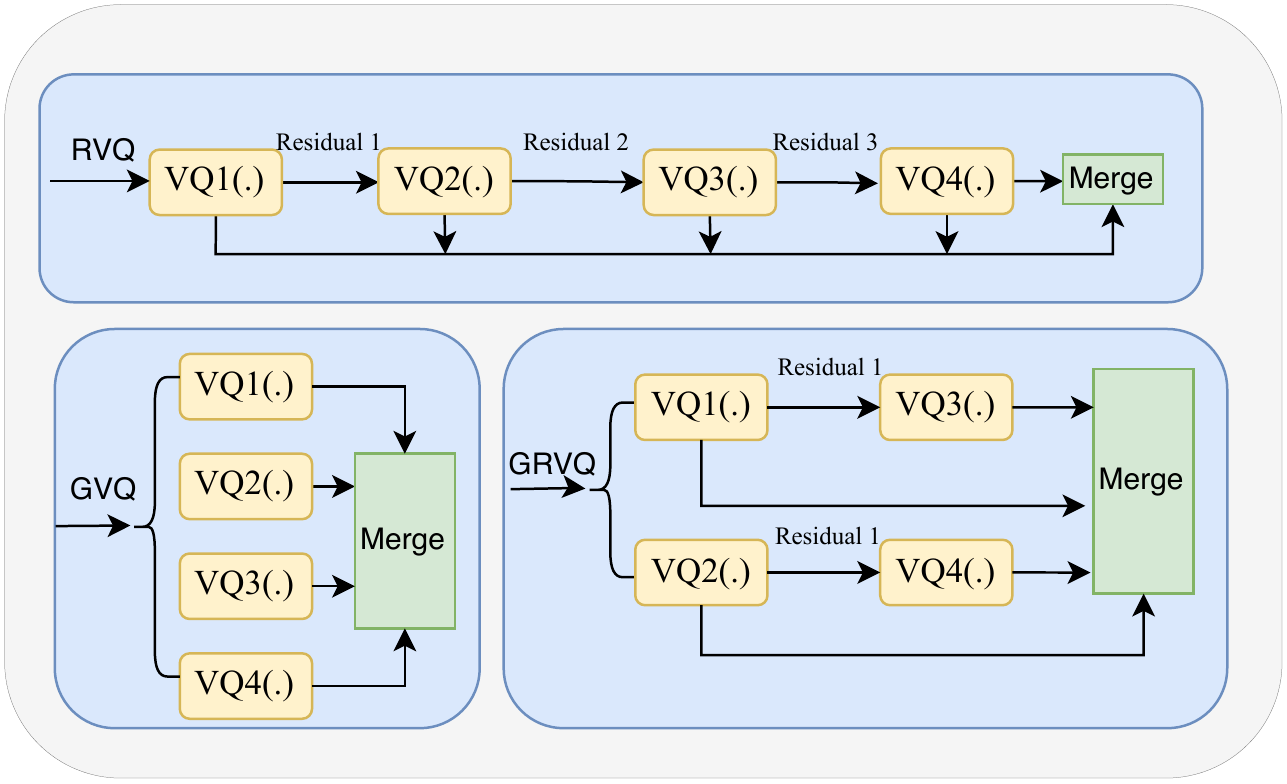}
  \caption{The three types of vector quantization. In this figure, we assume only using 4 vector quantization layer (4 codebooks).}
  \label{fig:vq-codec2}
  \vspace*{-\baselineskip}
\end{figure}

\subsubsection{Mel-VQ-Diffusion decoder}
With the help of the pre-trained Mel-VQ-VAE, we transfer the problem of mel-spectrogram prediction into that of predicting a sequence of quantization tokens. To generate high-quality mel-spectrogram tokens while maintaining fast inference speed, we propose a Mel-VQ-Diffusion decoder. In the following, we first introduce the basic idea of Mel-VQ-Diffusion, then summarize the training target. Lastly, we introduce classifier-free guidance to enhance the connection between conditional information and training targets. 

Given the paired training data $(\boldsymbol{x}_0, \boldsymbol{y})$, where $\boldsymbol{y}$ denotes the combination of phone features, style features and speaker features. $\boldsymbol{x}_0$ denotes the ground truth mel-spectrogram tokens. We first build a diffusion process, which corrupts the distribution of $p(\boldsymbol{x}_0)$ into a controllable stationary distribution $p(\boldsymbol{x}_T)$. Then we build a Transformer-based neural network \cite{vaswani2017attention} to learn to recover the $p(\boldsymbol{x}_0)$ conditioned on the $\boldsymbol{y}$. Inspired by previous works \cite{yang2022diffsound,gu2022vector}, we utilize a mask and uniform transition matrix to guide the diffusion process (we use the index $K+1$ denotes the mask token). The transition matrices $\boldsymbol{Q}_t \in \mathbb{R}^{(K+1) \times (K+1)}$ is defined as:
\begin{equation} \label{transition matrix}
 \boldsymbol{Q}_t =
 \begin{bmatrix}
    \alpha_t + \beta_t  & \beta_t & \beta_t & \beta_t & \cdots & 0 \\
    \beta_t &  \alpha_t + \beta_t & \beta_t  & \beta_t & \cdots & 0  \\
    \beta_t & \beta_t  & \alpha_t+\beta_t  & \beta_t & \cdots & 0  \\
    \vdots  &   \vdots     &  \vdots  &  \vdots     &  \ddots  &    \vdots \\
    \gamma_t & \gamma_t & \gamma_t & \gamma_t   &  \cdots & 1
\end{bmatrix}.
\end{equation}
The transition matrix denotes that each token has a probability of $\gamma_t$ transfers to the mask token, a probability of $K\beta_t$ be resampled uniformly over all K categories and a probability of $\alpha_t =  1-K\beta_t-\gamma_t$ to stay the original token. Based \cite{yang2022diffsound}, we can calculate $q(x_t|x_0)$ according to following formula:
\begin{equation}\label{formula:cal uniform and mask matrix}
   \overline{\boldsymbol{Q}}_t \boldsymbol{c}(x_0) = \overline{\alpha}_t \boldsymbol{c}(x_0) + ( \overline{\gamma}_t - \overline{\beta}_t )\boldsymbol{c}(K+1) + \overline{\beta}_t.
\end{equation}
where $\overline{\alpha}_T= \prod_{t=1}^{T} \alpha_t$, $\overline{\gamma}_T=1-\prod_{t=1}^{T}(1-\gamma_t)$ and $\overline{\beta}_T=(1-\overline{\alpha}_T - \overline{\gamma}_T)/K$. $\boldsymbol{c}(\cdot)$ denotes transfer a scalar element into a one-hot column vector. The stationary distribution $p(\boldsymbol{x}_T)$ can be:
\begin{equation}\label{formula:masked transition matrices}
  p(\boldsymbol{x}_T)=[\overline{\beta}_T, \overline{\beta}_T , \cdots, \overline{\gamma}_T],
\end{equation}
\textbf{Decoder Training Target}
We train a network $p_{\theta}(\boldsymbol{x}_{t-1}|\boldsymbol{x}_{t},\boldsymbol{y})$ to estimate the posterior transition distribution $q(\boldsymbol{x}_{t-1}|\boldsymbol{x}_t,\boldsymbol{x}_0)$. The network is trained to minimize the variational lower bound (VLB).
\begin{equation}\label{vlb loss}
\begin{aligned}
    \mathcal{L}_{\mathit{diff}} &= \sum_{t=1}^{T-1}  \big{[}D_{KL}[q(\boldsymbol{x}_{t-1}|\boldsymbol{x}_t,\boldsymbol{x}_0)||p_{\theta}(\boldsymbol{x}_{t-1}|\boldsymbol{x}_t,\boldsymbol{y})] \big{]} \\
    &+ D_{KL}(q(\boldsymbol{x}_T|\boldsymbol{x}_0)||p(\boldsymbol{x}_T)),
\end{aligned}
\end{equation}
\textbf{Enhancing the connection between $\boldsymbol{x}_0$ and $\boldsymbol{y}$} Based on previous discussion, we can see that the conditional information $\boldsymbol{y}$ inject into the network, to help optimize $p(\boldsymbol{x}_{t-1}| \boldsymbol{x}_t, \boldsymbol{y})$. However, in the last few steps, when $\boldsymbol{x}_t$ includes enough information, the network may ignore the conditional information $\boldsymbol{y}$ in the training stage. To solve this problem, we introduce the classifier free guidance \cite{ho2022classifier,tang2022improved} to enhance the connection between $\boldsymbol{x}_0$ and $\boldsymbol{y}$. Specifically, instead of only optimizing $p(\boldsymbol{x}|\boldsymbol{y})$, we expect to optimize the following target function:
\begin{equation} \label{formula:classifier-free}
  \log(p(\boldsymbol{x}|\boldsymbol{y})) + \lambda \log(p(\boldsymbol{y}|\boldsymbol{x})),
\end{equation}
where $\lambda$ is a hyper-parameter to control the degree
of posterior constraint.
Using Bayes's theorem, Formula (\ref{formula:classifier-free}) can be derived as:
\begin{equation}\label{formula:proof2}
\begin{aligned}
  \mathop{\arg\max}\limits_{x}[\log p(\boldsymbol{x}|\boldsymbol{y}) + \lambda \log p(\boldsymbol{y}|\boldsymbol{x})] \\
  = \mathop{\arg\max}\limits_{x}[(\lambda+1)\log p(x|y)-\lambda \log p(x)] \\
  = \mathop{\arg\max}\limits_{x}[\log p(x)+(\lambda+1)(\log p(x|y)-\log p(x))].
\end{aligned}
\end{equation}
To predict the unconditional mel-spectrogram token, we follow \cite{tang2022improved} to use a learnable null vector $\boldsymbol{n}$ to represent unconditional information $\boldsymbol{y}$. In the training stage, we set 10\% probability to use null vector $\boldsymbol{n}$. In the inference stage, we first generate the conditional mel-spectrogram token's logits $p_{\theta}(\boldsymbol{x}_{t−1} | \boldsymbol{x}_t, \boldsymbol{y})$, then predict the unconditional mel-spectrogram token's logits $p_\theta(\boldsymbol{x}_{t−1} | \boldsymbol{x}_t, \boldsymbol{n})$. Based on formula (\ref{formula:proof2}), the next step sample probability $p_{\theta}(\boldsymbol{x}_{t−1} | \boldsymbol{x}_t, \boldsymbol{y})$ can be re-write as:

\begin{equation} \label{formula:re-sample}
  p_{\theta}(\boldsymbol{x}_{t−1}|\boldsymbol{x}_t, \boldsymbol{n}) + (\lambda + 1)( p_{\theta}(\boldsymbol{x}_{t−1}|\boldsymbol{x}_t, \boldsymbol{y}) − p_{\theta}(\boldsymbol{x}_{t−1}|\boldsymbol{x}_t,\boldsymbol{n}))
\end{equation}

\begin{figure*}[t] 
  \centering
  \includegraphics[width=\linewidth]{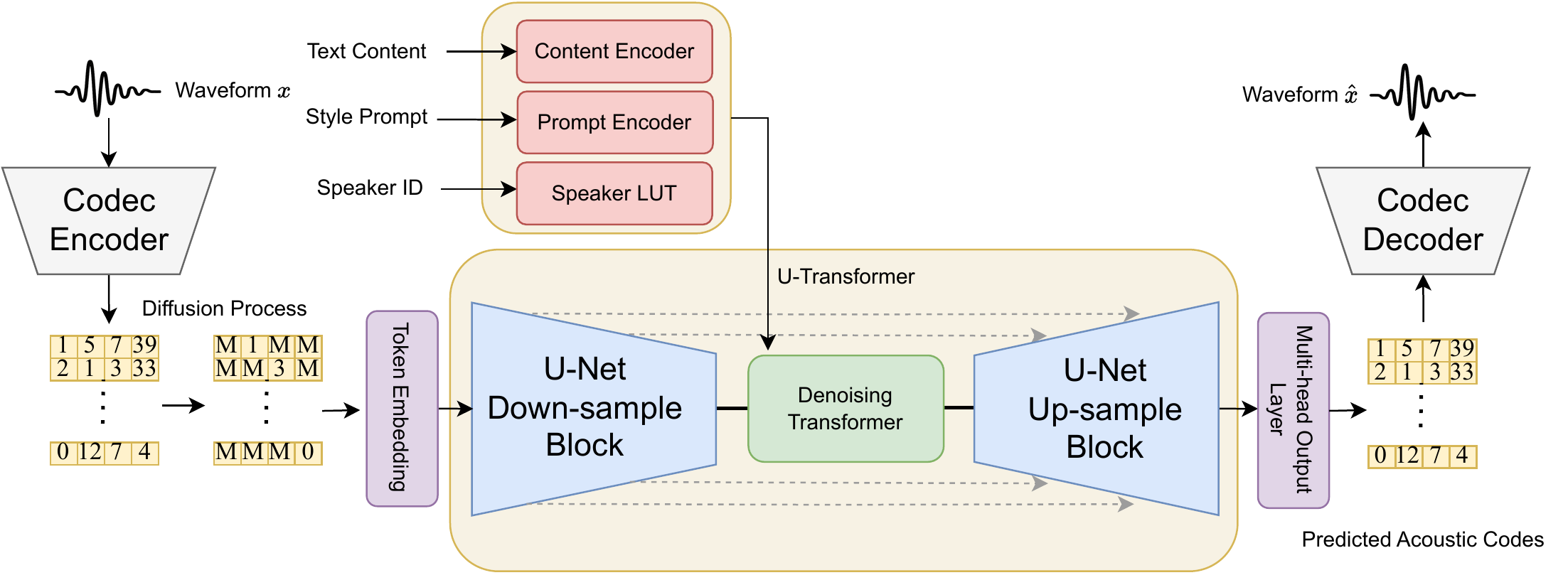}
  \caption{The framework of our proposed Wave-VQ-Diffusion. The U-transformer consists of several U-net down-sample and up-sample blocks and a denoising transformer.}
  \label{fig:wav-vq-diffusion}
  \vspace*{-\baselineskip}
\end{figure*}
\subsection{Modelling Waveform in Discrete Latent Space Via Multiple Vector Quantizers} \label{sec:residual-vector-q}
Inspired by the success of neural audio codec models, such as SoundStream \cite{zeghidour2021soundstream}, Encodec \cite{defossez2022high} and HiFi-Codec \cite{yang2023hifi}.
In this study, we additionally investigate directly predicting waveform in the discrete latent space with the help of large-scale pre-trained neural audio codec models. 

Recently, many methods have been proposed to generate speech using neural codec models, \textit{e.g.} AudioLM \cite{borsos2022audiolm} trains speech-to-speech language models on both k-means tokens from a self-supervised model and acoustic tokens from a neural codec model, leading to high-quality speech-to-speech generation. The concurrent work VALL-E \cite{wang2023neural} proposes to train a two-stage model to synthesize speech based on text input and reference audio. However, VALL-E needs a two-stage training strategy, and the first stage is an autoregressive language model, which has a significant influence on the synthesis speed.
In this study, we propose a non-autoregressive model based on the discrete diffusion model that significantly improves the synthesis speed while maintaining high-quality synthesis performance. 

As Figures \ref{fig:vq-codec} and \ref{fig:vq-codec2} show, compared to Mel-VQ-VAE, the neural audio codec model includes more codebooks. Although using more codebooks can improve reconstruction performance, it also raises a new research problem: How do you model such a long sequence by transformer? As we know, the computational complexity of a transformer is related to the sequence length. For a 10-second speech with 24k sampling rate, if we use 8 codebooks and set 240 times downsampling in the encoder, we will get 8000 tokens. Using a transformer based model to handle such a long sequence is challenging due to GPU memory limitations, thus it is necessary to seek a novel strategy for long sequence modelling. In this study, we propose a U-Transformer architecture to simultaneously model multiple codebooks.
As Figure \ref{fig:wav-vq-diffusion} shows, we first use several convolution layers to down-sample the input codebook matrix along the codebook number dimension. Then we use a denoising transformer to model the relationship of tokens in latent space. After that, we use several convolution layers and upsampling layers to recover the codebook number dimension. Lastly, we use different output layers to output prediction results for each codebook simultaneously.
\subsubsection{Wave-VQ-Diffusion}
There are three differences in Wave-VQ-Diffusion compared to Mel-VQ-Diffusion: (1) We adopt a U-transformer architecture to model multiple codebooks simultaneously. Note that we use the same transformer architecture as that in Mel-VQ-Diffusion. (2) We use different embedding tables for different codebooks due to the fact that tokens from different codebooks follow different data distributions. (3) We design an improved mask and uniform strategy for the diffusion process, which is based on the principle that the information included in the codebooks gradually decreases from the first residual vector quantization layer to the last layer. The codebooks in the first layer include most of the text, style, and speaker identity information. The following layers mainly include the fine-grained acoustic details, which are crucial for the speech's quality. We conjecture that the codebook's tokens in the first layer are easy to recover conditioned on $\boldsymbol{y}$, instead the following layer's tokens are hard to recover due to the fact that they have no obvious connection with $\boldsymbol{y}$.
Following the easy-first-generation principle, we should mask the last layer's codebook (\textit{e.g.} codebook $N_q$) at the start of the forward process and mask the foremost layer's codebook (\textit{e.g.} codebook 1) at the end of the forward process such that the learnable reverse process follows an easy-first generative behavior. However, previous commonly-used mask and uniform strategy assumed all of the tokens in the sequence were of the same importance, which violates the easy-first-generation principle. To solve this problem, we propose an improved mask and uniform strategy, whose details are presented in the following. \\
\textbf{Improved Mask and uniform strategy}
We dynamically allocate different weights for different codebooks when we pre-define the transition matrix. Considering these aforementioned properties, we construct $\overline{\alpha}_t^{i}$ , $\overline{\gamma}_t^{i}$ and $\overline{\beta}_t^{i}$ as follows
\begin{equation}\label{formula:alpha}
\begin{aligned}
  \overline{\alpha}_t^{i} = 1 - \frac{t}{T} - \frac{\exp(\frac{i\%N_q}{2*N_q})}{2*T}, \\
  \overline{\gamma}_t^{i} =  \frac{t}{T} + \frac{\exp(\frac{i\%N_q}{2*N_q})}{2*T}, \\
  \overline{\beta}_t^{i} = (1-  \overline{\alpha}_t^{i}-\overline{\gamma}_t^{i})/K,
\end{aligned}
\end{equation}
where $N_q$ denotes the index of residual layer in neural audio codec model, $i$ denotes the token position in the sequence. In our study, we concatenate all of the tokens from the first codebook to the last codebook.

\begin{algorithm}[t]
\caption{Training of the InstructTTS.}
\label{alg:PA1}
\begin{algorithmic}[1]
\REQUIRE ~~\\
    Pre-trained prompt encoder, A transition matrix $\boldsymbol{Q}_t$, timestep $T$, network parameters $\theta$, training epoch $N$, NLSpeech dataset $\boldsymbol{D}$, the encoder of VQ-VAE $E_{vq}$.
    \FOR{$i=1$ to $N$}
    \FOR{$(\mbox{conetent prompt}, \mbox{style prompt},  \mbox{audio})$ in $
    \boldsymbol{D}$}
    \STATE $mel = \mbox{get\_mel\_spectrogram(audio)}$;
    \STATE $\boldsymbol{x}_0 = E_{vq}(mel)$;
    \STATE $\boldsymbol{c}=$ContentEncoder($\mbox{content prompt}$);
    \STATE $\boldsymbol{z}_e = $AudioEncoder($\mbox{mel}$);
    \STATE $\boldsymbol{z}_p = $PromptEmb($\mbox{style prompt}$);
    \STATE $\boldsymbol{z}_s = $SpeakerEmb($\mbox{speaker id}$);
    \STATE $\boldsymbol{y}=\boldsymbol{c}+\boldsymbol{z}_e+\boldsymbol{z}_s$;
    \STATE sample $t$ from Uniform($1, 2, 3, ..., T$);
    \STATE sample $\boldsymbol{x}_t$ from $q(\boldsymbol{x}_t|\boldsymbol{x}_0)$ based on formula (\ref{formula: cal uniform and mask matrix});
    \STATE estimate $p_{\theta}(\boldsymbol{x}_{t-1}|\boldsymbol{x}_t,\boldsymbol{y})$;
    \STATE calculate loss according to formula (\ref{formula:training-func});
    \STATE update network $\theta$;
    \ENDFOR
    \ENDFOR
\RETURN network $\theta$.
\end{algorithmic}
\end{algorithm}

\begin{algorithm}[t]
\caption{Inference of the InstructTTS.}
\label{alg:PA2}
\begin{algorithmic}[1]
\REQUIRE ~~\\
    Time stride $\Delta_t$, timestep $T$, Content Prompt, Style Prompt, the decoder of VQ-VAE $G$, network $\theta$, stationary distribution $p(\boldsymbol{x}_T)$;
    \STATE $t=T$, $\boldsymbol{c}=$ContentEncoder($\mbox{content prompt}$);
    \STATE $\boldsymbol{z}_s = $SpeakerEmb($\mbox{speaker id}$);
    \STATE $\boldsymbol{z}_p = $PromptEmb($\mbox{style prompt}$);
    \STATE $\boldsymbol{y}=\boldsymbol{c}+\boldsymbol{z}_p+\boldsymbol{z}_s$;
    \STATE sample $\boldsymbol{x}_t$ from $p(\boldsymbol{x}_T)$;
    \WHILE{$t \textgreater 0$}
    \STATE sample $\boldsymbol{x}_t$ based on formula (\ref{formula:re-sample}) 
    \STATE $t \leftarrow (t-\Delta_t)$
    \ENDWHILE
\RETURN  $G(\boldsymbol{x}_t)$.
\end{algorithmic}
\end{algorithm}
\subsection{The Training and Inference Details}
In this section, we summarize the overall training objective and the inference process.
\subsubsection{Training objective}
Our proposed InstructTTS can be trained in an end-to-end manner. The overall training objective is as follows:
\begin{equation}\label{formula:training-func}
\begin{aligned}
  \mathcal{L} = \mathcal{L}_{\mathit{diff}} + \mathcal{L}_{\mathit{var}}+ \lambda_1 I(\boldsymbol{z}_e;c) + \lambda_2 I(\boldsymbol{z}_e;\boldsymbol{z}_{sid}) + \\
  \lambda_3 D_{Euc}(\boldsymbol{z}_p,\boldsymbol{z}_e) - \beta_1 \mathcal{F}_1(\theta_1) -  \beta_2 \mathcal{F}_2(\theta_2).
\end{aligned}
\end{equation}
where $\mathcal{L}_{\mathit{diff}}$ denotes the diffusion loss, $\mathcal{L}_{\mathit{var}}$ denotes the duration, pitch and energy reconstruction loss. $I(.)$ denotes mutual information, $D_{Euc}$ denotes the L2 loss. $\mathcal{F}_1 (\theta_1)$ and $\mathcal{F}_2 (\theta_2)$ denote the likelihood approximation model of $q_{\theta_1} (\boldsymbol{z}_{sid}|\boldsymbol{z}_e)$ and $q_{\theta_2} (\boldsymbol{z}_{e}|\boldsymbol{c})$ respectively. Details about the MI estimation and minimization can be found in \cite{cheng2020club}. 
The whole training process is summarized on Algorithm \ref{alg:PA1}. Note that we assume a Mel-VQ-Diffusion decoder is used in Algorithm 1. When we use a Wave-VQ-diffusion decoder, a similar process is used.
\subsubsection{Inference}
In the inference process, we directly use the feature extracted by style prompt embedding model as the style features. In our experiments, we set the $T=100$ and $\Delta_t=1$. The whole inference process is summarized on Algorithm \ref{alg:PA2}.

\section{Experimental setup}
\label{sec5:exp}
\subsection{Dataset and Data Pre-processing}\label{sec5:dataset}
\subsubsection{Dataset for Vector Quantization Pre-training}
To obtain a robust and acoustic-informative Vector Quantization model, we combine one internal dataset with the following three commonly-used public-available TTS datasets: (1) Our internal dataset, which is a Mandarin Chinese speech corpus, contains 300 hours speech data. (2) The VCTK dataset \footnote{\url{https://datashare.ed.ac.uk/handle/10283/2651}}. (3) The AISHELL3 dataset \cite{shi2020aishell}. (4) Clean splits of the LibriTTS dataset \cite{panayotov2015librispeech}. In total, the training set has 669 hours speech data.
\subsubsection{Dataset for InstructTTS} We use our internal dataset NLSpeech as our training and testing dataset. The details are presented in Section \ref{sec3:dataset}.
\subsubsection{Data pre-processing}
All audio clips have a sampling rate of 24kHz. For Mel-VQ-VAE pre-training, the log mel-spectrograms extracted using a 1024-points Hanning window with 240-points hop size and 80 mel bins. The PyWorld toolkit \footnote{\url{https://github.com/JeremyCCHsu/Python-Wrapper-for-World-Vocoder}} is used to compute F0 values from speech signals. Energy features are computed by taking the $l_2$-norm of frequency bins in STFT magnitudes.
\subsection{Implementation Details} \label{subsec:implementation details}
We first pre-train the Mel-VQ-VAE and neural audio codec models. Then we fix the pre-trained model, and train the InstructTTS model in an end-to-end manner. In the following, we will introduce the details of the network structure and training strategy.
\subsubsection{VQ-VAE} In this study, the network structure of the Mel-VQ-VAE model is similar to the VQ-GAN model \cite{esser2021taming,iashin2021taming}. To preserve more time-dimension information, we set a downsampling factor of 2 along the time axis, and a downsampling factor of 20 along the frequency axis. For the codebook $\boldsymbol{Z}$, the dimension of each code word vector $n_z$ is set as 256, and the codebook dictionary size $K$ is set as 512. In our experiments, we set the learning rate as $1\times 10^{-4}$. The Adam optimizer \cite{kingma2014adam} (the betas are 0.5 and 0.9) is adopted to optimize weights. 
\subsubsection{Neural Audio Codec Model} Inspired by the success of Encodec \cite{defossez2022high}, SoundStream \cite{zeghidour2021soundstream} and HiFi-Codec \cite{yang2023hifi}. We explore three types of audio codec models based on different quantization techniques: Residual Vector Quantization (RVQ) \cite{defossez2022high,zeghidour2021soundstream}, Group Vector Quantization (GVQ) \cite{chen2023vector} and Group Residual Vector Quantization (GRVQ) \cite{yang2023hifi}.
The RVQ-based audio codec model is much similar to the Encodec model except that the frame-shift is 10ms, resulting in 100 frames for 1-second audio.
The GVQ-based audio codec model follows the one in the MQTTS system \cite{mqtts}, which also uses a frame-shift of 10 ms.
The GRVQ-based audio codec model is the same as the one introduced in HiFi-Codec \cite{yang2023hifi} and we use the officially open-source implementation \footnote{\url{https://github.com/yangdongchao/AcademiCodec}}.
For RVQ model, we set the maximum codebook size as 12 in the training process. Similar to SoundStream \cite{zeghidour2021soundstream}, quantizer dropout is used. For the GVQ and GRVQ models, we set 4 codebooks in the training process. Each codebook includes 1024 code words for all of the models. We train the three types of audio codec models with the same training set as introduced in Section\ref{sec5:dataset}.
\subsubsection{InstructTTS} Our proposed InstructTTS consists of three main parts: a style encoder, a content encoder and a discrete diffusion decoder. For the content encoder, we follow FastSpeech2 \cite{ren2020fastspeech} and use the same architecture for the phoneme encoder and the variance adaptor. The style encoder contains a pre-trained prompt encoder model (the details presented in Section \ref{Prompt Embedding}) and an audio encoder. The audio encoder consists of two convolution layers and one multi-head attention module. For the discrete diffusion model, we follow a similar architecture as \cite{yang2022diffsound}, we build a 12-layer 8-head transformer with a dimension of 256 for the decoder. Each transformer block contains a full-context attention layer, a linear fusion layer to combine conditional features, and a feed-forward network block. For the default setting, we set timesteps $T = 100$. For the diffusion process, we adopt the linear schedule strategy, which linearly increases $\overline{\gamma}_t$ and $\overline{\beta}_t$ from 0 to 0.9 and 0.1, and decreases $\overline{\alpha}_t$ from 1 to 0. We optimize our network using the AdamW optimizer \cite{loshchilov2017decoupled} with $\beta_1 = 0.9$ and $\beta_2 = 0.94$. The basic learning rate is $3\times 10^{-6}$, and the batch size is 16 for each GPU.
\subsection{Baseline Approach}

In the literature, there is no existing expressive TTS model using natural language style prompt to control stylish generation. Following the traditional neural TTS paradigm which predicts intermediate acoustic features (e.g., mel-spectrograms) from text input, we adapt the StyleSpeech model proposed in \cite{min2021meta} as the baseline approach. We replace the Mel-Style-Encoder in the StyleSpeech model with the same style encoder module used in InstructTTS, making the comparison as fair as possible. The baseline model uses the same HiFi-GAN vocoder to generate waveform as the proposed model using Mel-VQ-VAE.

\section{Evaluation Metric} \label{sec6:eval}
\subsection{Objective Evaluation}
We evaluate the synthesized speech from two aspects: speech quality and prosody similarity. For speech quality, we adopt Mel-cepstral distorion (MCD) \cite{kubichek1993mel}, structural similarity index measure (SSIM) \cite{wang2004image} and Short-Time Objective Intelligibility (STOI) \cite{taal2010short} to evaluate the speech quality. For prosody similarity, we use three pitch-related metrics: Gross Pitch Error (GPE), Voicing Decision Error (VDE) \cite{nakatani2008method} and F0 Frame Error (FFE) \cite{chu2009reducing}. GPE, VDE and FFE are widely applied to evaluate the performance of expressive TTS. The details of these metrics will be introduced as follows. 
\subsubsection{Mel-cepstral distorion} Spectral features, based on the short-term power spectrum of sound, such as Mel-cepstral coefficients (MCEPs), contain rich information about expressivity and emotion \cite{bitouk2010class}. Mel-cepstral Distortion (MCD) \cite{kubichek1993mel} is a widely adopted metric to measure the spectrum similarity, which is computed as
\begin{equation}\label{formula:mcd}
\begin{aligned}
  MCD = \frac{1}{T} \sum_{t=0}^{T-1} \sqrt{\sum\limits_{m=1}^M (c_{m,t}-\hat{c}_{m,t})^2},
\end{aligned}
\end{equation}
where $c_{m,t}$ and $\hat{c}_{m,t}$ denote the $m$-th mel-frequency cepstral coefficient (MFCC) of the $t$-th frame from the reference
and synthesized speech. We sum the squared differences over the first $M$ MFCCs. In this study, we set $M=24$.
\subsubsection{SSIM and STOI} Structural similarity index measure (SSIM) \cite{wang2004image} and Short-Time Objective Intelligibility (STOI) \cite{taal2010short} are effective metrics to evaluate the speech clarity and intelligibility. Following previous work \cite{liu2022diffgan}, we also adopt them as one of the metrics for speech quality.
\subsubsection{Prosody-related metrics} Given that pitch is considered as a major prosodic factor contributing to speech emotion and closely correlated to the activity level \cite{johnson1986recognition,owren2007measuring}, in this study, we adopt GPE, VDE and FFE as 
pitch similarity metrics to evaluate the synthesis results.
The metrics of GPE, VDE and FFE have been used as common objective evaluation metric for expressive TTS \cite{skerry2018towards}.
\begin{table*}[t] \centering
\caption{Objective and subjective evaluation as well as model size results. MCD, SSIM, STOI, GPE, VDE and FFE  are adopted as objective metrics. GT denotes the ground truth speech, GT (voc) denotes that we use pre-trained vocoder (HiFi-GAN) recover speech from mel-spectrogram. MOS and RMOS are presented with 95\% confidence intervals.}
\label{tab:exp1}
\begin{tabular}{cccccccccc}
\hline
\textbf{Model}               & \textbf{Decoder}  & \textbf{MCD}($\downarrow$) & \textbf{SSIM}($\uparrow$) & \textbf{STOI} ($\uparrow$) & \textbf{GPE}($\downarrow$) & \textbf{VDE}($\downarrow$) & \textbf{FFE}($\downarrow$)  & \textbf{MOS}($\uparrow$) & \textbf{RMOS}($\uparrow$) \\ \hline
GT   & \multicolumn{7}{c}{-}  & 4.62 $\pm$ 0.05         & 4.65 $\pm$ 0.05   \\ 
GT (voc)  & -   & 5.02        & 0.695         & 0.893         & 0.006       & 0.076       & 0.08                   &  4.41 $\pm$ 0.07        & 4.61 $\pm$ 0.07          \\ \hline
Baseline  & Mel-decoder       & 5.75        & 0.422          & 0.663         & 0.433        & 0.286       & 0.33        & 4.04 $\pm$ 0.08        & 3.85 $\pm$ 0.1          \\ \hline
\multirow{4}{*}{InstructTTS} & Mel-VQ-Diff  & \textbf{5.59}       & \textbf{0.487}  & \textbf{0.732}   & 0.392  & 0.246 & 0.30    & \textbf{4.35 $\pm$ 0.07}   & 4.22 $\pm$ 0.09          \\
& Wave-VQ-Diff (GVQ) & 5.85      & 0.356         & 0.564         & 0.384    & 0.193        & 0.27      & 3.44 $\pm$ 0.07   & {4.08 $\pm$ 0.06}  \\ 
& Wave-VQ-Diff (RVQ) & 5.77      & 0.365         & 0.587         & 0.370    & 0.166        & 0.25       & 3.59 $\pm$ 0.08   & {4.27 $\pm$ 0.07}  \\
& Wave-VQ-Diff (GRVQ) & 5.68      & 0.384         & 0.615         & \textbf{0.359}    & \textbf{0.151}        & \textbf{0.23}       & 3.95 $\pm$ 0.05   & \textbf{4.32 $\pm$ 0.07} \\ \hline
\end{tabular}
\end{table*}
\subsection{Subjective Evaluation}
To further validate the effectiveness of our proposed method, we conduct subjective evaluation from two aspects: speech quality and style relevance. 
\subsubsection{Speech quality}
We first conduct the Mean Opinion Score (MOS) test to
evaluate speech quality, which aims to evaluate the speech’s naturalness, fidelity and intelligibility. All participants are asked to listen to the reference speech (“Ground truth”) and the
synthesized speech and score the “quality” of each speech sample on a 5-point scale (‘5’ for excellent, ‘4’ for good, ‘3’ for fair, ‘2’ for poor, and ‘1’ for bad). Each audio sample is rated by at least 20 testers. 
\subsubsection{The style's relevance between synthesized speech and the natural language prompt}
We conduct RMOS (relevance mean opinion score) for speaking style relevance on the testing set to evaluate the relevance between synthesized speech and the prompt. All participants are asked to read the natural language prompt and then listen to the synthesized speech. After that, the participants are asked to score the “relevance” of each speech sample on a 5-point scale (‘5’ for excellent, ‘4’ for good, ‘3’ for fair, ‘2’ for poor, and ‘1’ for bad). Each audio sample is rated by at least 20 testers.
\subsubsection{AXY test}
We propose to use AXY test \cite{skerry2018towards} to assess the style relevance between the generated speech with its corresponding natural language style prompt. Raters are asked to rate a 7-point score (from -3 to 3) and choose the speech samples which sound closer to the natural language style prompt in terms of style expression. For each  natural language style prompt (A), the listeners are asked to choose a preferred one among the samples synthesized by the baseline model (X) and proposed Method (Y), from which AXY preference rates are calculated. The score less than 0 represents that “X is much closer", and the score more than 0 represents that “Y is much closer". Note that we do not use the ground truth speech as reference, instead we ask raters to read the natural language style prompt, and then evaluate which synthesized speech is closer to the prompt in terms of semantic meaning in emotion and style.

\begin{table}[t] \centering
\caption{The AXY preference test results for speaking style relevance.}
\label{tab:AXY-test}
\begin{tabular}{ccc}
\hline
X                         & Y                  & 7-point score \\ \hline
\multirow{2}{*}{Baseline} & InstructTTS (Mel)  &    0.72           \\
                          & InstructTTS (Wave) &     0.84          \\ \hline
\end{tabular}
\end{table}

\begin{table}[t] \centering
\caption{The emotion classification probability (\%) comparison between our proposed methods and the baseline. For each type of emotion, we choose 15 samples. The table reports the averaged probability values of 15 utterances.}
\label{tab:ecp}
\begin{tabular}{ccccc}
\hline
Model             & Sad   & Happy & Angry & Overall \\ \hline
GT                & 100   & 88.80  & 94.70  & 95.20    \\ \hline
Baseline          & 64.28 & 66.60  & 68.15 & 66.70    \\
InstructTTS (Mel) & \textbf{71.42} & \textbf{66.60}  & 68.40  & 69.10    \\
InstructTTS (Wave) & \textbf{71.42} & 55.50  & \textbf{84.21} & \textbf{71.42}   \\ \hline
\end{tabular}
\end{table}

\subsection{Emotion Perception Test} \label{Emotion Perception Test}
Given that speaking style is related to emotion. We choose three types of test samples (happy, sad and angry) from our test set based on the natural language prompt. Then we expect our proposed methods can generate similar emotional speech with the guidance of natural language prompt. We propose to use emotion classification probability to validate the emotion perception performance. Intuitively, the classification probabilities summarize the useful emotion information from the previous layers for final output layer. Thus, we believe that the classification probabilities can be an effective tool to justify the synthesized speech's performance. To realize this, we first pre-train an emotion classification model in our internal emotion classification dataset. We adopt a pre-trained wav2vec2 \cite{baevski2020wav2vec} model as feature extractor, and then we add two linear layers and one softmax layer.
\section{RESULTS AND ANALYSIS}\label{sec7:result}
In this section, we conduct experiments to verify the effectiveness of our proposed InstructTTS. We first compare the performance between our proposed InstructTTS and the baseline. Then we conduct ablation studies to validate the effectiveness of each part of our proposed methods. 
\begin{table}[t] \centering
\caption{The ablation study for cross-modal representation learning. We evaluated with the test set of Chinese STS-B corpus. SCC denotes Spearman Correlation Coefficient.}
\label{tab:exp2}
\begin{tabular}{cc}
\hline
Model                    & SCC (\%) \\ \hline
w/o cross-modal learning & 80.4                                  \\ \hline
w cross-modal learning   & \textbf{80.94}                                 \\ \hline
\end{tabular}
\end{table}
\subsection{The comparison between InstructTTS and the Baseline}
\subsubsection{The analysis of objective metrics}
Table \ref{tab:exp1} shows the objective metrics (MCD, SSIM, STOI, GPE, VDE, FFE) comparison between our proposed InstructTTS and the baseline system. We have the following observations: (1) Our proposed InstructTTS achieves better performance than the baseline system in terms of speech quality and prosody. (2) Using Mel-VQ-Diffusion as decoder can realize better speech quality than Wave-VQ-Diffusion, but Wave-VQ-Diffusion is superior in maintaining prosody details. One of the reasons is that the pre-trained Mel-VQ-VAE downsamples 20 times along the frequency dimension, which may harm the pitch information.
Instead, Wave-VQ-Diffusion directly models all of the information in time domain, prosody-related information can be well reserved, but some acoustic details may sacrifice. The concurrent work VALL-E \cite{wang2023neural} also faces the same problem that speech quality is suboptimal.
We find that the audio codec significantly influences the speech quality. As the last three lines in Table \ref{tab:exp1} shows, different audio codec modelling can bring different synthesis performance. We can see that using GRVQ model brings the best performance in Wave-VQ-Diffusion. In the following, without specifically stated, we use GRVQ by default for Wave-VQ-Diffusion model.
\subsubsection{Subjective Evaluation}
We conduct crowd-sourced mean opinion score (MOS) tests to evaluate the quality of the synthesized speech perceptually. Furthermore, we also conduct crowd-sourced relevance mean opinion score (RMOS) tests to evaluate the relevance between the synthesized speech and the prompt. The results are shown on Table \ref{tab:exp1}. We can see that InstructTTS (mel) gets the best MOS performance, and InstructTTS (wave) gets the best RMOS performance. We can see that both the two types of our proposed InstructTTS models obtain better RMOS performance than the baseline. The subjective evaluation results are consistency with the objective evaluation results. We can also observe that the speech quality of InstructTTS (wave) still has room for improvement in quality, on which we will further study in our future work.

We additionally conduct AXY preference test to compare InstructTTS and the baseline in terms of the naturalness of prosody in their generated speech. From Table~\ref{tab:AXY-test}, we can see that the raters show much higher preference to the proposed InstructTTS (Mel) and InstructTTS (Wave) than to the baseline model.

\subsubsection{Emotion Perception Evaluation}
To further evaluate the expressiveness in modeling speaking emotion and styles with InstructTTS, we conduct perception evaluation with a speech emotion classification model. 
The details are introduced in Section \ref{Emotion Perception Test}. The results are reported in Table \ref{tab:ecp}. We can see that the pre-trained speech emotion classification (SEC) model obtains good classification performance in the ground truth set, which proves that our SEC model is effective. Furthermore, we can observe that InstructTTS gets better classification performance than the baseline, with the InstructTTS (Wave) model getting the best performance. We note that the evaluation results are consistent with the FFE results.  \\
\begin{table}[t] \centering
\caption{The text-to-audio retrieval performance in the test set. We use Recall at rank k (R@k) as the metrics.}
\label{tab:exp3}
\begin{tabular}{cccc}
\hline
Loss Type        & { R@1} & {R@5} & {R@10} \\ \hline
Contrastive Loss & 11.62                            & 42.97                            & 61.72                             \\ \hline
InfoNCE          & \textbf{15.62}                            & \textbf{42.97}                            & \textbf{63.67}                             \\ \hline
\end{tabular}
\end{table}

\begin{table}[t] \centering
\caption{The ablation study for the effectiveness of classifier-free guidance (CFG) and mutual information minimization (MIM) training strategy.}
\label{tab:exp4}
\begin{tabular}{ccccccc}
\hline
Model  & MIM & CFG & MCD($\downarrow$) & SSIM($\uparrow$)  & FFE($\downarrow$) \\ \hline
\multirow{4}{*}{InstructTTS (Mel)} & \usym{2613} &\usym{2613} & 5.75 & 0.421 &0.42 \\
& \usym{2613} &\checkmark & 5.65 & 0.442 & 0.37 \\
& \checkmark &\usym{2613} & 5.66 & 0.451 &0.32 \\
& \checkmark &  \checkmark  & \textbf{5.59}    & \textbf{0.487}         & \textbf{0.30}     \\ \hline
\multirow{4}{*}{InstructTTS (Wav)} & \usym{2613} &\usym{2613}   & 5.85 & 0.350  & 0.39    \\
&\usym{2613} &\checkmark   & 5.74 & 0.365  & 0.34    \\
&\checkmark &\usym{2613}   & 5.75 & 0.353  & 0.33    \\
&\checkmark &\checkmark  & \textbf{5.68}    & \textbf{0.384}          & \textbf{0.23}    \\ \hline
\end{tabular}
\end{table}
\subsection{Ablation studies for InstructTTS}
\subsubsection{The impact of cross-modal representation learning for robust style embedding}
In this section, we explore the effectiveness of our proposed cross-modal representation learning in Section \ref{Prompt Embedding}. Table \ref{tab:exp2} presents the results. We can see that, after finetuning with our proposed cross-modal representation learning, the performance of the RoBERTa even achieves better performance in STS task than only finetuning with SimCSE. Furthermore, we also evaluate the text-to-audio retrieval performance in the test set. As Table \ref{tab:exp3} shows, we can see that using InfoNCE loss as training objective can bring better retrieval performance than the contrastive ranking loss.
\subsubsection{The Impact of Mutual information minimization (MIM) training} In this study, we propose to using mutual information minimization strategy to constrain the encoded information by the audio encoder, we expect the audio encoder only encodes the style-related information. In this part, we conduct ablation studies to investigate whether our proposed MIM strategy can bring better performance. The experiments results report on Table \ref{tab:exp4}. We can see that using MIM training strategy can improvement in both speech quality and pitch similarity (especially
 in pitch similarity), which proved that effectiveness of feature disentangled strategy. 
\subsubsection{The effectiveness of classifier-free guidance} 
In this study, we propose to use classifier-free guidance (CFG) strategy to enhance the connection between conditional information and the predicted results. To validate the effectiveness of classifier-free guidance, we conduct ablation study, the experiments are shown on Table \ref{tab:exp4}. We can see that using CFG strategy can bring better performance due to it enhancing the connection between conditional information and the predicted results, which forces the model to better utilize the conditional information.
\subsubsection{The effectiveness of improved diffusion strategy}
Table \ref{tab:exp5} shows the experiments results when we use different diffusion strategies. We can see that our proposed improved mask and uniform strategy can bring better performance. The experimental results validate our proposed easy-first-generation principle.
\begin{table}[t] \centering
\caption{The ablation study for different diffusion strategy. MAR represents the mask and replace strategy. I-MAR denotes our improved mask and replace strategy. Note that we conduct experiments on Wave-VQ-Diffusion-based InstructTTS.}
\label{tab:exp5}
\begin{tabular}{ccccc}
\hline
Model               & MCD($\downarrow$) & SSIM($\uparrow$) & STOI($\uparrow$) & FFE($\downarrow$) \\ \hline
InstructTTS (MAR)   &  5.71   &  0.364    & 0.582     &  0.24   \\ \hline
InstructTTS (I-MAR) &  \textbf{5.68}   &  \textbf{0.384}    & \textbf{0.615}     &  \textbf{0.23}   \\ \hline
\end{tabular}
\end{table}
\begin{table}[t] \centering
\caption{The neural audio codec's reconstruction  performance comparison. $N_q$ denotes we use $N_q$ codebooks to reconstruction the audio.}
\label{tab:exp6}
\begin{tabular}{ccccc}
\hline
Model                 & $N_q$ & Down-sample times   & PESQ   & STOI   \\ \hline
\multirow{3}{*}{RVQ (ours)} & 2  &240    & 2.63  & 0.87  \\
                      & 4   &240  & 3.24 & 0.91 \\
                      & 8   &240     & 3.54   & 0.93  \\ \hline
GVQ (ours) & 4  &240  & 3.11    & 0.91  \\ \hline
GRVQ (ours) & 4 &240  &  \textbf{3.63}     & \textbf{0.95}  \\ \hline
Encodec (Facebook)  & 12 &320    & 3.21  & \textbf{0.95}  \\ \hline
\end{tabular}
\end{table}
\subsubsection{Exploring the influence of audio codec for speech synthesis.}
As we discuss in Section \ref{subsec:implementation details}, we train three types of audio codec models by using different vector quantization techniques: Residual Vector Quantization (RVQ), Group Vector Quantization (GVQ) and Group Residual Vector Quantization (GRVQ). For GVQ and GRVQ, we use 4 codebooks in training and inference process. For RVQ, we use 12 codebooks in the training stage. To fair compare with the GVQ and GRVQ, we also choose 4 codebooks in the inference stage. We use an out-of-domain test set (including 1024 high-quality 24kHz audio samples) to evaluate the reconstruction performance of three types of neural audio codec model and the pre-trained Encodec models \cite{defossez2022high}. Table \ref{tab:exp6} shows the reconstruction performance (we adopt the popular metrics PESQ and STOI from speech enhancement fields to measure the reconstruction performance). We can see that the GRVQ models obtain the best reconstruction performance. Table \ref{tab:exp2} also shows that using the GRVQ model can synthesis better speech, which means that the audio codec models significantly influence the performance of generation model. We can see that using more codebooks (\textit{e.g.} using 8 codebooks in RVQ models) can improve reconstruction performance. However, we find that the performance of the generated models is not proportional to the number of codebooks. We think one of the reason is that using more codebooks will bring burden for generation models. Furthermore, we conjecture that the amount of data in the NLSpeech dataset is still in-sufficient and using a larger-scale dataset can bring extra improvement.
\begin{figure}[t] 
  \centering
  \includegraphics[width=\linewidth]{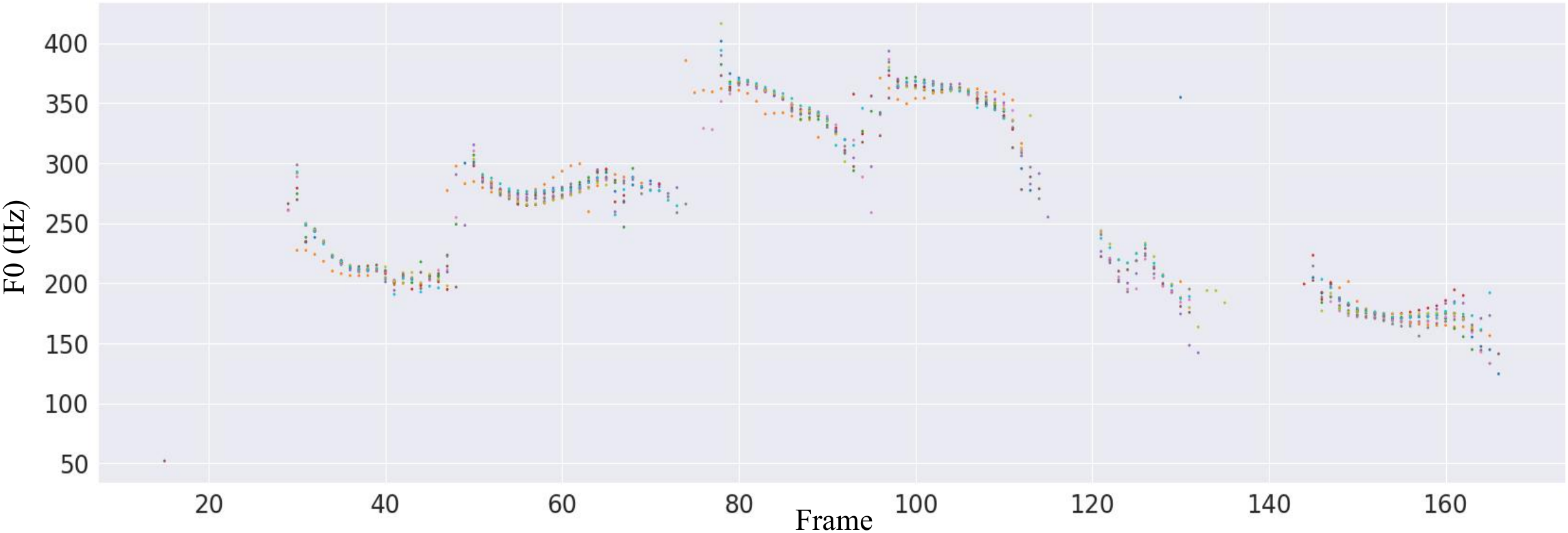}
  \caption{Pitch tracks. We present the F0 contours of 10 different runs with the same text input, speaker id and style prompt conditioning.}
  \label{fig:vis}
  \vspace*{-\baselineskip}
\end{figure} 
\subsection{Synthesis Variation}
Unlike the baseline systhem, which
output is uniquely determined by the input text and other conditional information (such as speaker identity, natural language prompt) at inference, InstructTTS takes sampling
processes at denoising steps and can inject some variations
into the generated speech. To demonstrate this, we run
a InstructTTS (mel) model 10 times for a particular
input text, speaker and natural language prompt, and then compute the F0 contours of the generated speech samples. We visualize in Figure \ref{fig:vis} and observe that InstructTTS can synthesize speech with diverse pitches.

\section{Conclusion}
\label{sec8:conclusion}
In this work, we present InstructTTS, which can synthesize expressive speech with the natural language prompt. To our best of knowledge, this is the first work to use long and complex natural language prompt to control the speaking style. In terms of acoustic model, we propose a novel perspective to model expressive TTS: we propose to model expressive TTS in the discrete latent space and cast speech synthesis as a language modeling task. We explore two kinds of modelling methods: (1) modelling mel-spectrogram with the help of a pre-trained Mel-VQ-VAE model; (2) modeling waveform with the help of a pre-trained neural audio codec model. In terms of model structure, we propose a novel U-transformer, which can effectively model long-sequence. Our experiments demonstrate the advantages of our proposed method. 

This work still has some limitations that need to be addressed in our future work: (1) The inference speed is limited due to the diffusion step is large (we use 100 diffusion steps). (2) We will build large-scale dataset to train the InstructTTS models, similar to VALL-E and AudioLM. We believe that InstructTTS is expected to be more robust when the amount of training data increases.

\section*{Acknowledgments}
We thank the help of our colleagues Mingjie Jin and Dan Su for this paper. They help us build the NLSpeech dataset. 
\bibliographystyle{ieeetr}
\balance
\bibliography{refs.bib}
\end{document}